\documentclass[a4paper, 11pt,onecolumn,final,notitlepage,oneside]{article}
\usepackage[cp1251]{inputenc}
\usepackage[T1,T2A]{fontenc}
\usepackage[english]{babel}
\usepackage{indentfirst}
\usepackage{geometry}
\usepackage{graphicx}
\usepackage[final]{epsfig}
\usepackage{multicol}
\usepackage{graphicx}
\usepackage{subfigure}
\usepackage{amsmath}
\usepackage{enumerate}
\usepackage{amssymb}
\usepackage{amscd}
\usepackage{hhline}
\usepackage{multirow}
\usepackage{setspace}
\usepackage{makeidx}
\usepackage{textcomp}
\usepackage{amsfonts}
\usepackage{afterpage}
\usepackage{longtable}
\usepackage{cite}
\usepackage{rawfonts}
\usepackage{array}
\usepackage{amsopn}
\usepackage{hyperref}
\hypersetup{colorlinks=true, linkcolor=blue, citecolor=green, filecolor=magenta, urlcolor=magenta, pdfpagemode=FullScreen}

 \graphicspath{{pictures/}}

\DeclareGraphicsExtensions{.pdf,.png,.jpg,.eps}

\makeatletter
\def\@seccntformat#1{\csname the#1\endcsname.\ } 
\def\@biblabel#1{#1.} 
\makeatother

\begin{document}

\title{\normalsize \begin{flushleft}
{UDC 531.011, 531.314.6, 531.395}
\end{flushleft}
\vspace{\baselineskip}
 \normalsize \bf Jacobi’s principle of stationary full action and its implications}
\author{\small Vitaliy V. Voytik,\,\\
\small \itshape Cand. Phys. \& Math. Sci., Associate Professor\\  \small \itshape Dep. of Medical Physics and Informatics, Bashkir State Medical University, \\ \small \itshape 3, Lenin street, Ufa, 450008, Russian Federation\\
\small \itshape e-mail: vvvojtik1@bashgmu.ru\\}
\date{}
\maketitle
\renewcommand{\abstractname}{}

\begin{abstract}
The aim of this study is to extend the scope of Jacobi’s stationary fullaction principle to nonconservative natural systems and to derive the equations of motion consistent with this extended principle.

To this end, in addition to the conventional variation of Jacobi’s action
with respect to coordinates, an independent variation with respect to time is
introduced. The small variations of coordinates and time are treated as functions of a point on the true trajectory and are required to satisfy standard
vanishing boundary conditions at the endpoints.

The variation of Jacobi’s action with respect to time yields a geometric
form of the well-known theorem on the change of kinetic energy for a natural
system. The coordinate variation leads to a second-order equation governing
the curvature of the trajectory as a function of the tangent vector, external
fields, and intrinsic properties of the system.

These equations, together with a first-order differential relation determining the speed of motion along the configuration-space trajectory, form
a system of $s+1$ second-order equations. Its solution describes the natural
form of the law of motion, i.e. it fixes both the trajectory and the law of
motion along it as functions of arc length. The solution depends on $2s$ initial
constants: the initial kinetic energy, the initial coordinates, and the initial
tangent vector.

From the three original equations, the Lagrange equations for natural
systems are derived. It is also shown that the variation operator with respect to coordinates and time is invariant across different formulations of
the stationary-action principle. This implies that the three forms of the
principle—the modified Hamilton action, the Hamilton–Ostrogradsky action, and Jacobi’s action—are equivalent. Consequently, Jacobi’s method
is applicable to nonconservative mechanical systems. Thus, the extended
Jacobi principle provides the basis for an independent and equivalent formulation of analytical mechanics.
\end{abstract}

\newpage
\begin{flushleft}
{\bf{Introduction}}
\end{flushleft}

In the calculus of variations, it is essential that variations in new variables correspond to variations in the original ones. Otherwise, one might arrive at different conditions for the extremum of a functional. An example of the application of variations is the analytical mechanics of holonomic systems, which is based on a specific integral functional—the action. Unless the Hamilton–Jacobi method is employed, each representation of the action in terms of new variables corresponds to its own distinct variational approach.

There are three known forms of the principle of stationary action: the Hamilton – Ostrogradsky principle, the modified Hamilton principle, and the Jacobi principle (sometimes referred to as the Maupertuis principle). Each form of the action is associated with its own specific variational conditions and equations of motion. The combination of these properties defines a particular formulation of mechanics. The first two principles are straightforward, whereas the Jacobi principle is more difficult to grasp.
  
The background to the development of this principle prior to Jacobi, along with the original works of its founders, can be found in the monograph \cite{18} and the collection of papers \cite{34}, respectively. Initially, Jacobi’s principle of stationary total action was presented in a reduced form \cite[lectures 6, 7]{1}. The significance of this principle in mechanics was highlighted by Hertz \cite{77}. Historically, the principle in its Jacobi form is closely related to the less useful principle of Lagrange. Jacobi’s principle itself is often undervalued; for instance, while its formulation has remained consistent for a long time, it does not appear in every textbook. Nevertheless, the modern understanding of Jacobi’s principle is covered in a variety of manuals and textbooks—though not always in great detail—for example, \cite[p. 3.8]{44}, \cite[p. 45, Г]{8},  \cite[ch. 4]{37}, \cite[p. 2.3, 7.4.4]{19}, \cite[ch. VII, p. 4, s/p 2, 3]{23}, \cite[p. 2.7]{16}, \cite[p. 7.2.2.]{11}, \cite[p. 26]{29}, \cite[p. 6.3, 6.4]{36}, \cite[p. 20, 21]{32}, \cite[p. 8.6]{10},  \cite[p. 8.12]{7}, \cite[p. 9.2.3]{20}, \cite[p. 44]{15}, \cite[ch. 5, p. 6, 7]{17}, \cite[p. 4-6]{13},  \cite[p. 12.10, 12.11]{26}, \cite[ch. XIII, p. 2]{31}, \cite[p. 8.1]{12}, \cite[app. 5]{63}, \cite[p. 2.3.4]{6}, \cite[p. 9.10]{28},  \cite[ch. IV, p. 17, 18]{21},  \cite[ch. VI, p. 5, 6]{3}, \cite[p. 202]{2}. This principle is used to derive the Jacobi equation for the trajectory of a system in space \cite[p. 20, 21]{32}, \cite[p. 12.10, (12.11.6)]{26}. In this process, not all coordinates are varied; one of them remains fixed, as it plays the role of time \cite[lecture 6]{1}, \cite[(12.11.4)]{26}, \cite[p. 193, (32.69)]{2}. This is considered necessary for obtaining the equations of the trajectory, and it renders them non-covariant. 
 
In contemporary scientific literature, the Jacobi principle enables the use of topological methods to identify new examples of integrable geodesic flows \cite{46} and periodic trajectories \cite[ch. 4]{37}. This principle allows for a transition from one Hamiltonian system to another that is trajectory-equivalent to it—a fact with far-reaching implications. The reader may find recent results in this field, as well as in related areas, in references \cite{48}–\cite{56}.

The relevance of this article stems from a long-standing problem in the foundations of mechanics concerning the various forms of the action functional and their variations.

First and foremost, it should be emphasized that all three types of action are considered equal and mutually equivalent. However, for non-conservative systems, only two of them are applicable, as the Jacobi action is not used in such cases (see \cite{44}-\cite{2}). Thus, the scope of the Jacobi action is significantly narrower compared to other types of action that do not require a system to be conservative. This limitation renders the Jacobi action less comprehensive than other formulations.

When deriving equations of motion from the principle of stationary action, an additional issue arises: because the variations of coordinates and time—as well as the boundary conditions—differ across the various types of action, the variation operator itself changes. In the Hamilton–Ostrogradsky action (and its modifications), the boundaries for coordinate and time variations are fixed, while the variations of both quantities are arbitrary and independent. In the Lagrange action—which is related to the Jacobi action—the boundary conditions require constant energy (isoenergetic conditions); here, the time variations at the end of the path are non-zero, which couples them to the coordinate variations. In the Jacobi action, time variations are not used; for non-conservative systems, they are considered to be devoid of meaning.
This renders the variations used in the Lagrange and Jacobi actions non-equivalent to those used in the Hamilton action.

These weaknesses in the theory may have been an additional reason for the remark cited in \cite[footnote to item 45, G]{8}, as they hinder a complete understanding of Jacobi’s principle and its application to non-conservative systems. Although these shortcomings have been known for a long time, to the best of the author’s knowledge, no significant attempts have been made to date to rectify the situation.

The identified problems can be resolved in a fundamental way. Since the action, in any of its forms, is essentially equivalent to the others, one need only account for the time dependence of the system’s energy and its characteristic parameters—namely, external fields and the system’s intrinsic tensor parameter. In this way, the action becomes applicable to the general case, including non-conservative systems. In other words, the constraint $E = \text{const}$ — typically imposed on the Jacobi action—is superfluous and should be removed. For any formulation of the complete action, the variation process must involve the same number of coordinate and time variations, which must remain arbitrary and independent, subject to consistent boundary conditions. Only under these circumstances can one assert the mathematical equivalence of the variation operator (with respect to coordinates and time) across different forms of the action. This premise implies that the theorem of the stationarity of the complete Jacobi action holds not only for variations with respect to coordinates but also for variations with respect to time. The present work is devoted, in part, to substantiating this claim.

The principle under discussion is of significant interest. According to the authors \cite[Ch. 4, Sec. 10]{35}, the calculus of variations serves as an exceptionally reliable tool for deriving and investigating the differential equations of mathematical physics. Consequently, the equations obtained through this principle will help uncover robust, general patterns that underpin the mechanics of non-conservative systems within a trajectory-based framework.
Furthermore, this approach provides a theoretical foundation for studying natural systems in variable fields by employing a geometric, intuitive representation—a feature of great importance for practical applications.

The object of study is the mechanics of non-conservative natural systems, based on the *natural method of describing motion*. In this context, the "natural" specification of a system's motion entails defining the generalized \textit{hodograph} — the trajectory $q^\alpha = q^\alpha(q)$ — and the time function $t = t(q)$ (the \textit{schedule}), both expressed in terms of the arc length of the trajectory \cite[sections 37, 41]{2}, \cite[ch. 3, pp. 1.3, 2.1]{33}. While this approach is commonly employed in the kinematics of a material point, this paper applies it to the dynamics of a natural system. The subject of the study is a modified principle of stationary total action in the Jacobi form, along with its implications.

The formulation of the Jacobi principle employs the condition of natural parametrization for the system's law of motion. This assumption will be justified in Section 7, following the proof of the principle. The proof relies on a result from \cite{30}. In that work, two equations for the trajectory-based method—one governing the rate of change of kinetic energy and the other defining the trajectory itself—were rigorously derived from the general Lagrange equations for a natural system. These relations must be connected to the modified Jacobi principle. It is necessary to verify the validity of this assertion. If taking the variations of the Jacobi action with respect to the coordinates and time yields these specific equations, then the theorem concerning the stationarity of the Jacobi action under such variations will be proven.

In this case, to obtain generally covariant equations of the trajectory (which is important for their solution), all coordinates are varied equally, as shown  in \cite[Section 2.7]{16}, \cite[problem for Section 44]{15}, \cite[formula (12.12.22)]{26}, and \cite[example following (9.251)]{28}. Another example of the derivation of similar equations is provided in \cite[problem for Section 87]{25}.

In this regard, the following objectives are addressed sequentially in this work. First, a modified version of Jacobi’s principle is formulated.
Next, it is proven that the residual and reduced components of the total Jacobi action are stationary; this serves as the basis for deriving the general equations of motion as consequences of the principle. Subsequently, the Lagrange equations are derived from these results. Following this, a comparison is made between the variations of the action—interpreted as functions—within the frameworks of Jacobi and Hamilton–Ostrogradsky. Finally, it is shown that the equations of trajectory mechanics effectively define the natural form of motion in configuration space, which helps clarify the meaning of the variations in the Jacobi action. This sequence aligns with the logic of the subsequent exposition. As a preliminary step, Section 1 provides introductory information regarding the Jacobi action.

The purpose of this article is to demonstrate the need to extend the scope of Jacobi’s principle of least (stationary) action to non-conservative systems.

\subsubsection{Full, shortened, and residual actions}

In the Lagrangian formulation, the mechanical system moves in such a way that the quantity
\begin{equation}\label{1.1}
S=\int\limits^{t_2}_{t_1}L(q^\alpha, \dot{q}^\alpha, t)dt\,,
\end{equation}
remains stationary under small variations of the coordinates and time. The action \eqref{1.1} is known as the \textit{Hamilton-Ostrogradsky action}.

If a mechanical system is \textit{natural}, its Lagrangian has the form 
\begin{equation}\label{1.2}
	L=\frac{1}{2}\,\mu_{\alpha\beta}\dot{q}^{\alpha}\dot{q}^{\beta}+P_{\alpha}\dot{q}^{\alpha}-U\,.
\end{equation}
Conservative mechanical systems with such a Lagrangian are sometimes referred to as generalized \cite[p. 11, 20]{32} or quadratic \cite[ch. 3, p. 6]{4} systems. We shall assume that the coefficients appearing in this Lagrangian are arbitrary functions of the coordinates and time. This form of the Lagrangian function is characteristic of any mechanical system described within the framework of classical mechanics. Examples include a deformable body with a moment of inertia tensor $\mu_{\alpha\beta}$ or a point particle in a Riemannian space with a metric tensor $\mu_{\alpha\beta}$. Here, $U$ is termed the potential energy, and $P_\alpha$ is called the potential momentum (a term introduced by Ch. Kittel \cite[appl. G]{22})).
Potential energy typically describes the interaction of a system with an electric or gravitational field, whereas potential momentum accounts for the interaction with a magnetic field.

Let's derive the Jacobi action using \eqref{1.2}.
To do this, let us calculate the energy $E$ of the natural system as a function of coordinates, velocities and time. Let's first find $p_\alpha$:
\[	p_\alpha=\frac{\partial L}{\partial \dot{q}^\alpha}=\mu_{\alpha\beta}\dot{q}^\beta+P_\alpha\,.\]
Hence,
\begin{equation}\label{1.3}
	E=p_\alpha \dot{q}^\alpha-L=\frac{1}{2}\,\mu_{\alpha\beta}\dot{q}^\alpha \dot{q}^\beta+U\,.
\end{equation}

Let us substitute the definition of the generalized velocity, $\dot{q}^\alpha = dq^\alpha/dt$, into this equation and express $dt$. We obtain 
\begin{equation}\label{1.10}
	2T=\frac{\mu_{\alpha\beta} dq^\alpha dq^\beta}{dt^2}=\left(\frac{dq}{dt}\right)^2\,,
\end{equation}
or
\begin{equation}\label{1.4}
	dt=\frac{dq}{n}\,,
\end{equation}
where the quantity $n$
\begin{equation}\label{1.5}
	n=\sqrt{2T}\,,
\end{equation}
is called the refractive index, $T=E-U$ is the kinetic energy, and
\[		dq=\sqrt{\mu_{\alpha\beta}\,dq^\alpha dq^\beta}\]
is a line element between nearby points in a Riemannian \textit{configuration} space with a metric $\mu_{\alpha\beta}$.

Let us express the potential energy in this equation in terms of the total and kinetic energies and substitute it into \eqref{1.1}, taking \eqref{1.3} into account. We obtain
\[	S=\int\left(\mu_{\alpha\beta}\,\dot{q}^{\alpha}\dot{q}^{\beta}+P_{\alpha}\dot{q}^{\alpha}-E\right)dt\,.\]
We replace integration over time with integration along the trajectory. This change of variables involves the use of the energy equation \eqref{1.3}, and the action then takes the following form \cite[Corollary 8.12.2]{7}.
\begin{equation}\label{1.7}
	S=\int \left(\sqrt{2(E-U)}\,\,dq+P_{\alpha}d{q}^{\alpha}\right)-\int Edt\,.
\end{equation}
 Let’s represent all the functionality as
\begin{equation}\label{1.8}
	S=\int \limits^{q_2}_{q_1}\left(ndq+P_{\alpha}d{q}^{\alpha}\right)-\int\limits^{q_2}_{q_1} \left(\frac{n^2}{2}+U\right)dt\,.
\end{equation}
The first integral in formulas \eqref{1.7} and \eqref{1.8} is called Jacobi’s \textit{shortened action}. The second integral in these equations will henceforth be referred to as the \textit{residual action}, while the entire action is termed the \textit{complete action} in Jacobi’s form. In the absence of a magnetic external field, the shortened action formally coincides with the optical path length of a ray as defined in Fermat’s principle. The significance of the integral \eqref{1.8} will be clarified later.

\subsubsection{Jacobi's theorem on the stationarity of the complete action}

Let us consider the possible variations within the Jacobi action. The standard variations in \eqref{1.7} are those taken with respect to the coordinates. If a natural system is conservative, the coordinates are the only variables that appear in the action. In the general case—that is, for non-conservative systems—the situation changes. External fields (such as the refractive index and the potential momentum), as well as the defining characteristic of a natural system—the metric tensor of the configuration space—may also depend on time. This gives rise to a question: should the Jacobi action also be varied with respect to time?

To answer this question, we note that in the Hamilton-Ostrogradsky action \eqref{1.1} and the modified Hamilton action, time can be varied as an independent variable. In the Jacobi action,  time enters the functional via a schedule $t = t(q)$; therefore, varying with respect to time implies varying this function. Consequently, by virtue of the equivalence of these action formulations, time must also be varied in the Jacobi action, albeit subject to the aforementioned parametrization.

\begin{figure}[ht]
\noindent
\begin{minipage}[h]{0.48\textwidth}
\caption{Projection of the trajectory and the general vector of coordinate variation onto the surface  $(q^1q^2)$ 
\label{image1} }
\end{minipage} \hfill
\begin{minipage}[h]{0.5\textwidth}
\centering
\includegraphics[width=1.0\textwidth]{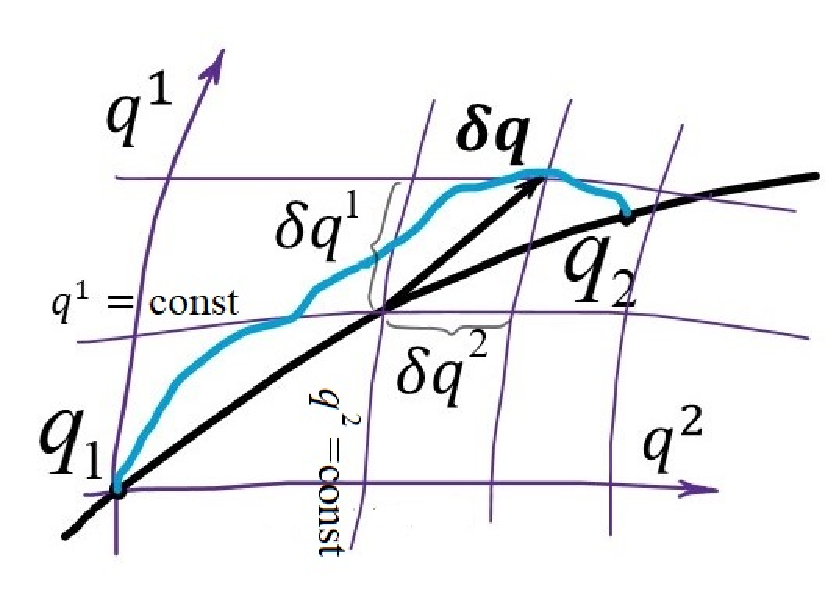}
\end{minipage}
\end{figure}

Let us assume that Jacobi’s principle asserts the minimality of the action for a natural description of the true motion of a system.
Let’s explore the meaning of the variations in full action. Let $q^\alpha = q^\alpha(q)$ and $t = t(q)$ be the actual trajectory (hodograph) of the system and its schedule — that is, the functions for which the action \eqref{1.8} attains a stationary value.This means that the Jacobi action remains unchanged under the substitution $q^\alpha(q) \rightarrow q^\alpha(q) + \delta q^\alpha(q)$, where the function $\delta q^\alpha(q)$ is arbitrary and small along the entire trajectory. In this context, the vector $\boldsymbol{\delta q}$ itself defines a correspondence between a point on the true trajectory (shown in black) and the corresponding point on the varied trajectory (see Figure \ref{image1}) (shown in light blue).

A similar variation of the action with respect to time implies that the Jacobi action remains unchanged under the substitution $t(q) \rightarrow t(q) + \delta t(q)$, where the function $\delta t(q)$ is also small along the entire trajectory. 

 Given that the solution to the equations of motion admits a natural parametrization, it is clear that the integrands in the Jacobi action depend solely on the path length $q$. Furthermore, the integration in the Jacobi action \eqref{1.8} is evidently performed with respect to this variable, from $q_1$ to $q_2$.

Naturally, the explicit form of the variations in the Jacobi action differs somewhat from those in the Hamilton-Ostrogradsky action; in the latter, the action remains invariant under the substitution $q^\alpha(t) \rightarrow q^\alpha(t) + \delta q^\alpha(t)$, where the function $\delta q^\alpha(t)$ is small over the entire time interval from the beginning to the end of the motion. Nevertheless, in both actions \eqref{1.1} and \eqref{1.8}, the variations of time and coordinates are arbitrary and independent within their respective parameterizations (in \eqref{1.1}, they are functions of time $t$; in \eqref{1.8}, they are functions of the label $q$).

Thus, the total action in the Jacobi form is stationary with respect to variations of the coordinates and time under natural parametrization. In other words,
the extended [principle] holds. In other words, an \texttt{extended theorem regarding the stationarity of the total Jacobi action} holds true.

\textit{Let a natural system move along a certain generalized hodograph $q^\alpha = q^\alpha(q)$ according to a schedule $t = t(q)$ between an initial point and a final point, which are specified by the initial label $q_1$ and the final label $q_2$ on this curve. Then, the total action in the form \eqref{1.8} attains a stationary value, $\delta S = 0$, under arbitrary and independent variations of the coordinate functions $q^\alpha(q)$ and the time function $t(q)$ at the initial and final points of the trajectory—corresponding to the labels $q_1$ and $q_2$—provided that the boundary conditions $\delta q^\alpha(q_1) = \delta q^\alpha(q_2) = \delta t(q_1) = \delta t(q_2) = 0$ are satisfied.}

Let us now consider how a mechanical system must move in order for the action in the Jacobi form to remain invariant under simultaneous small variations of the coordinates and time. In the calculations that follow, we shall consistently omit the dependence of the variations on $q$.

\subsubsection{Stationarity of the residual action.}

During the motion of a non-conservative system, its total energy changes with time: $T + U = E(t)$.
Taking the partial derivative of this equation with respect to a coordinate leads to the relation
		\begin{equation}\label{3.13}
		\frac{\partial T}{\partial q^\alpha}=-\frac{\partial U}{\partial q^\alpha}
	\end{equation}
 or
\begin{equation}\label{3.14}
		n\,\frac{\partial n}{\partial q^\alpha}+\frac{\partial U}{\partial q^\alpha}=0\,.
	\end{equation}

Let us discuss the variation of the residual action, taking into account equality \eqref{3.14}.

First, note that the term involving the total energy $E$ in the residual action is an integral over time. When varying this term with respect to time, the resulting boundary term, $\left(n^2/2+U\right)\delta t\big|^{q_2}_{q_1}$, vanishes due to the boundary conditions $\delta t(q_1)=\delta t(q_2)=0$ imposed in the theorem regarding the stationarity of Jacobi’s total action. This reasoning can be confirmed by a more detailed calculation.

The variation of the residual action is equal to
	\[\delta \check{S}=-\int \left(\delta T dt+\delta Udt +T d\delta t+Ud\delta t\right)\,.\]
The variation of the first term is of the form
\[	\delta Tdt=\delta \left(\frac{n^2}{2}\right)dt=\left(n\,\frac{\partial n}{\partial t}\,\delta t + n\,\frac{\partial n}{\partial q^\alpha}\,\delta q^\alpha\right)dt\,.\]
The second term is equal to
	\[\delta Udt=\frac{\partial U}{\partial q^\alpha}\delta q^\alpha dt+\frac{\partial U}{\partial t}\delta t dt\,.\]
We will integrate the third and fourth terms by parts:
	\[T d\delta t=d(T\delta t)-dT \delta t=d(T\delta t)-ndn \delta t=d(T\delta t)-\left(n\frac{\partial n}{\partial q^\alpha}dq^\alpha+n\frac{\partial n}{\partial t}dt\right) \delta t\,,\]
\[	Ud\delta t=d(U\delta t)-dU \delta t=d(U\delta t)-\left(\frac{\partial U}{\partial t}\,dt+\frac{\partial U}{\partial q^\alpha}\,dq^\alpha\right) \delta t\,.\]
	
Summing these equations, we note that the terms involving $\delta t dt$ cancel each other out. Grouping the remaining terms associated with $\delta q^\alpha$ and $\delta t$, we finally obtain
\begin{equation}\label{3.25}
	\delta \check{S}=-\left(\frac{n^2}{2}+U\right)\delta t\bigg|^{q_2}_{q_1}-\int\limits^{q_2}_{q_1} \left(n\,\frac{\partial n}{\partial q^\alpha}+\frac{\partial U}{\partial q^\alpha}\right)\left(\delta q^\alpha dt - \delta t dq^\alpha\right)\,.
\end{equation}
  	The integrated term vanishes by virtue of the boundary conditions. Furthermore, in view of \eqref{3.14}, the integrand $n\, \partial n/\partial q^\alpha+\partial U/\partial q^\alpha$ is identically zero; therefore, the integral also vanishes. Consequently, the residual action is stationary.

		\subsubsection{Stationarity of the shortened action.}	
			
		Let us now determine the result of varying the reduced action $\tilde{S}$ with respect to the coordinates and time
				\begin{equation}\label{4.14}
	\delta \tilde{S}=\delta \int\limits^{q_2}_{q_1}\left(ndq+P_{\alpha}d{q}^{\alpha}\right)\,.
\end{equation}

\textbf{4.1. Calculation of the variation.} By varying the integrand in \eqref{4.14} with respect to the coordinates $q^\alpha$ and time $t$, while keeping the limits of integration fixed, we obtain
\begin{equation}\label{4.2}
	\delta \tilde{S}=\int\limits^{q_2}_{q_1}\left(\delta n dq+n\delta dq+\delta P_\alpha dq^\alpha+P_\alpha d\delta q^\alpha\right)=0\,. 
\end{equation}
Let us consider sequentially each term in the integrand \eqref{4.2}.

The first term containing the variation of the refractive index $\delta n$ has the form
\[	\delta n dq=\left(\frac{\partial n}{\partial q^\gamma}\,\delta q^\gamma+\frac{\partial n}{\partial t}\,\delta t\right) dq\,.\]

The second term is related to the variation of the length element $dq$. By varying the square of the length, we have
\[\delta dq^2=2dq\,\delta dq=\delta\left(\mu_{\alpha\beta}dq^\alpha dq^\beta\right)=\left(\frac{\partial\mu_{\alpha\beta}}{\partial t}\,\delta t+\frac{\partial \mu_{\alpha\beta}}{\partial q^\gamma}\,\delta q^\gamma\right)\tau^\alpha \tau^\beta dq^2+\]
\[	+2\mu_{\gamma\beta}\tau^\beta dq\, d\delta q^\gamma \,.\]   
 It follows from this
\[\delta dq=\frac{1}{2}\left(\frac{\partial\mu_{\alpha\beta}}{\partial t}\delta t+\frac{\partial \mu_{\alpha\beta}}{\partial q^\gamma}\delta q^\gamma\right)\tau^\alpha \tau^\beta dq+\mu_{\gamma\beta}\tau^\beta  d\delta q^\gamma \,.\]  
Multiplying this equation by $n$ and integrating the second term by parts, we obtain
\[	n\delta dq=\frac{n}{2}\left(\frac{\partial\mu_{\alpha\beta}}{\partial t}\delta t+\frac{\partial \mu_{\alpha\beta}}{\partial q^\gamma}\delta q^\gamma\right)\tau^\alpha \tau^\beta dq+d\left(n\mu_{\gamma\beta}\tau^\beta\delta q^\gamma\right)-d\left(n\mu_{\gamma\beta} \tau^\beta\right) \delta q^\gamma\,.\]

Let us examine the last differential in more detail and use the equality \eqref{1.4}:
\[	d\left(n\mu_{\gamma\beta}\,\tau^\beta\right)=n\mu_{\gamma\beta}\,\frac{d\tau^\beta}{dq}\, dq+\frac{\partial\left(n\mu_{\gamma\beta}\right)}{\partial t} \,\tau^\beta \frac{dq}{n}+ \frac{\partial\left(n\mu_{\gamma\beta}\right)}{\partial q^\alpha} \,\tau^\alpha\tau^\beta dq\,.\]
By substituting and expanding the partial derivatives, we obtain an expression for the second term in the form
\[n\delta dq=\frac{n}{2}\left(\frac{\partial\mu_{\alpha\beta}}{\partial t}\,\delta t+\frac{\partial \mu_{\alpha\beta}}{\partial q^\gamma}\,\delta q^\gamma\right)\tau^\alpha \tau^\beta dq+d\left(n\mu_{\gamma\beta}\tau^\beta\delta q^\gamma\right)-n\mu_{\gamma\beta}\frac{d\tau^\beta}{dq}\,\delta q^\gamma dq-\]
\[	-\tau^\beta \left(\frac{\mu_{\gamma\beta}}{n}  \frac{\partial n}{\partial t}+ \frac{\partial\mu_{\gamma\beta}}{\partial t}  \right)\delta q^\gamma dq-\left(\mu_{\gamma\beta}\,\tau^\alpha\tau^\beta\frac{\partial n}{\partial q^\alpha}+n\,\frac{\partial\mu_{\gamma\beta}}{\partial q^\alpha}\,\tau^\alpha\tau^\beta\right) \delta q^\gamma dq\,.\]

The third term of the variation is of the form:
\[	\delta P_\alpha dq^\alpha=\frac{\partial P_\alpha}{\partial t}\,\tau^\alpha \delta t\, dq+\frac{\partial P_\alpha}{\partial q^\gamma}\,\tau^\alpha \delta q^\gamma dq\,.\]

The fourth term —
		\[		P_\alpha d\delta q^\alpha=d\left(P_\alpha \delta q^\alpha\right)-dP_\gamma\, \delta q^\gamma=d\left(P_\alpha \delta q^\alpha\right)-\left(\frac{\partial P_\gamma}{\partial t}\,dt+\frac{\partial P_\gamma}{\partial q^\beta}\, dq^\beta    \right)\delta q^\gamma\,.\]
Again using the equality $dt=dq/n$, we rewrite it as
\[P_\alpha d \delta q^\alpha = d (P_\alpha \delta q^\alpha) - \left( \frac{1}{n} \frac{\partial P_\gamma}{\partial t} + \frac{\partial P_\gamma}{\partial q^\beta}\, \tau^\beta \right) \delta q^\gamma dq\,.\]

Substituting all the components into \eqref{4.2} and grouping the terms involving $\delta t$ and $\delta q^\gamma$, we obtain the variation of Jacobi’s reduced action with respect to the coordinates and time in the form
\[	\delta \tilde{S}=\left[\left(n\mu_{\beta\gamma} \tau^\beta+P_\gamma\right)\delta q^\gamma\right]\bigg|^{q_2}_{q_1}+\int\limits_{q_1}^{q_2} \left[ \frac{\partial n}{{\partial t}} \right. + \frac{\partial P_\alpha}{\partial t}\,\tau ^\alpha  + \frac{n}{2}\left. \frac{\partial \mu _{\alpha \beta }}{\partial t}\,\tau ^\alpha \tau ^\beta  \right]\delta t\,dq +\]
			\[+\int\limits_{q_1}^{q_2} \left[ \frac{\partial n}{\partial q^\gamma} - n\mu_{\beta \gamma }\frac{d\tau ^\beta}{dq}   \right.- \mu_{\beta \gamma}\frac{\partial n}{\partial q^\alpha }\,\tau ^\alpha \tau ^\beta - \left( {\frac{{{\mu _{\beta \gamma }}}}{n}\frac{{\partial n}}{{\partial t}} + \frac{\partial \mu _{\beta \gamma }}{\partial t} + \frac{\partial {P_\gamma }}{{\partial {q^\beta }}} - \frac{\partial P_\beta}{\partial q^\gamma}} \right){\tau ^\beta }-\]
			\begin{equation}\label{4.8}
		- \frac{1}{n}\frac{\partial P_\gamma}{\partial t} + \frac{n}{2}\left( {\frac{{\partial {\mu _{\alpha \beta }}}}{{\partial {q^\gamma }}} - \frac{{\partial {\mu _{\gamma \beta }}}}{{\partial {q^\alpha }}} - \frac{{\partial {\mu _{\alpha \gamma }}}}{{\partial {q^\beta }}}} \right)\tau ^\alpha \tau ^\beta \biggr]\,\delta {q^\gamma }dq  = 0\,.
\end{equation}

\textbf{4.2. The law of change in kinetic energy.} The first term resulting from integration in the variation of the action vanishes at the boundaries of the trajectory.
Given the arbitrary nature of the time variation $\delta t$ in the first integral, we set the integrand equal to zero. Consequently,
\begin{equation} \label{5.12}
	\frac{\partial n}{\partial t}=-\frac{\partial P_\alpha}{\partial t}\,\tau ^\alpha  - \frac{n}{2} \frac{\partial \mu _{\alpha \beta }}{\partial t}\,\tau ^\alpha \tau ^\beta\,.
\end{equation}
Let us multiply this equation by the refractive index $n$, bringing it inside the differential.
Then, by substituting the value of $n$ from \eqref{1.5} into equation \eqref{5.12}, the latter can be rewritten as follows:
\begin{equation} \label{5.13}
	\frac{\partial T}{\partial t}=-\sqrt{2T}\,\,\frac{\partial P_\alpha}{\partial t}\,\tau ^\alpha  - T \,\frac{\partial \mu _{\alpha \beta }}{\partial t}\,\tau ^\alpha \tau ^\beta\,.
\end{equation}
Let us calculate the total derivative $dT/dt$. Since kinetic energy is a function of coordinates and time, we can write that
\[	\frac{dT}{dt}=\frac{\partial T}{\partial t}+\frac{\partial T}{\partial q^\alpha}\,\frac{dq^\alpha}{dt}\,.\]
Substituting \eqref{3.13} and \eqref{5.13} into this expression and taking into account that $dq^\alpha/dt=\sqrt{2T}\,\tau^\alpha$, we obtain the equation 
	\begin{equation} \label{2.21}
	\frac{dT}{dt}=\sqrt{2T}\left(-\frac{\partial U }{\partial q^\alpha}-\frac{\partial P_\alpha}{\partial t}\right)\tau ^\alpha  - T \,\frac{\partial \mu _{\alpha \beta }}{\partial t}\,\tau ^\alpha \tau ^\beta\,.
\end{equation}
This equality was presented in \cite[formula (3.3)]{30}.

		\textbf{4.3. Equation of the trajectory.} In the second integral, due to the arbitrariness of the coordinate variations $\delta q_\gamma$, the integrand is also equal to zero.
Therefore,
	\[ n\mu_{\beta \gamma }\frac{d\tau ^\beta}{dq} =\frac{\partial n}{\partial q^\gamma} - \mu_{\beta \gamma}\frac{\partial n}{\partial q^\alpha }\,\tau ^\alpha \tau ^\beta - \left( {\frac{{{\mu _{\beta \gamma }}}}{n}\frac{{\partial n}}{{\partial t}} + \frac{\partial \mu _{\beta \gamma }}{\partial t} + \frac{\partial {P_\gamma }}{{\partial {q^\beta }}} - \frac{\partial P_\beta}{\partial q^\gamma}} \right)\tau ^\beta -\]
	\begin{equation}\label{6.17}
	-\frac{1}{n}\frac{\partial P_\gamma}{\partial t} +\frac{n}{2}\left( {\frac{{\partial {\mu _{\alpha \beta }}}}{{\partial {q^\gamma }}} - \frac{{\partial {\mu _{\gamma \beta }}}}{{\partial {q^\alpha }}} - \frac{{\partial {\mu _{\alpha \gamma }}}}{{\partial {q^\beta }}}} \right)\tau ^\alpha \tau ^\beta\,.
\end{equation}
We substitute the expression for $\partial n/\partial t$ from \eqref{5.12} into equation \eqref{6.17}, multiply the resulting equation by $n$, and substitute the definition of the refractive index from \eqref{1.5}. As a result, we obtain the equation 
\[	2T\mu_{\gamma\beta}\,\frac{d\tau^{\beta}}{dq}=-\frac{\partial U}{\partial q^{\gamma}}-\frac{\partial P_{\gamma}}{\partial t}+\sqrt{2T}\left(\frac{\partial P_{\alpha}}{\partial q^{\gamma}}-\frac{\partial P_{\gamma}}{\partial q^{\alpha}}-\frac{\partial \mu_{\gamma\alpha}}{\partial t}\right)\,\tau^{\alpha}+\]
	\[	+\left[T\left(\frac{\partial\mu_{\alpha\beta}}{\partial q^{\gamma}}-\frac{\partial\mu_{\gamma\beta}}{\partial q^{\alpha}}-\frac{\partial\mu_{\gamma\alpha}}{\partial q^{\beta}}\right)+\mu_{\gamma\beta}\left(\frac{\partial U}{\partial q^{\alpha}}+\frac{\partial P_{\alpha}}{\partial t}\right)\right]\,\tau^{\alpha} \tau^{\beta}+\]
	\begin{equation}\label{2.20}
	+\frac{\sqrt{2T}}{2}\mu_{\gamma\beta}\,\frac{\partial \mu_{\epsilon\alpha}}{\partial t}\,\tau^{\alpha} \tau^{\beta}\tau^{\epsilon}\,.
\end{equation}
This equation was also considered in \cite[(4.2)]{30}.

\textbf{4.4. Variation of the Jacobi action as a function.} Let us now consider the variation of the total Jacobi action, evaluated for solutions to the equations of motion (i.e., when the integral terms in \eqref{3.25} and \eqref{4.8} vanish). In this case, the variation of the total action reduces to a sum of boundary terms
\begin{equation}\label{6.21}
	\delta S=\delta \check{S}+\delta \tilde{S}=\left[\left(n\mu_{\alpha\beta} \tau^\beta+P_\gamma\right)\delta q^\alpha-\left(\frac{n^2}{2}+U\right)\delta t\right]\bigg|^{q_2}_{q_1}\,.
\end{equation}

	\subsubsection{Derivation of Lagrange’s equations} 
	
	We shall show that Lagrange’s equations follow from the equalities \eqref{1.4}, \eqref{2.21}, and \eqref{2.20}. To this end, we introduce the definition of generalized acceleration in the Jacobi formulation:
	\[\ddot{q}^{\,\beta}=\frac{d}{dt}\left(\sqrt{2T}\,\tau^\beta\right)=\frac{1}{\sqrt{2T}}\frac{dT}{dt}\,\tau^\beta+\sqrt{2T}\,\frac{d\tau^\beta}{dt}\,.\]
		Substituting \eqref{2.21} and using \eqref{1.4}, we obtain
\[	\ddot{q}^{\,\beta}=\left(-\frac{\partial U }{\partial q^\alpha}-\frac{\partial P_\alpha}{\partial t}\right)\tau ^\alpha\,\tau^\beta-\frac{\sqrt{2T}}{2} \,\frac{\partial \mu _{\alpha \gamma}}{\partial t}\,\tau ^\alpha \tau ^\beta\,\tau ^\gamma+2T\,\frac{d\tau^\beta}{dq}\,.\]
Let us multiply this expression by $\mu_{\gamma\beta}$, use equation \eqref{2.20}, and apply \eqref{1.4} once again. We obtain the equality
\[\mu_{\gamma\beta}	\ddot{q}^{\,\beta}=-\frac{\partial U}{\partial q^{\gamma}}-\frac{\partial P_{\gamma}}{\partial t}+\left(\frac{\partial P_{\alpha}}{\partial q^{\gamma}}-\frac{\partial P_{\gamma}}{\partial q^{\alpha}}-\frac{\partial \mu_{\gamma\alpha}}{\partial t}\right)\,\dot{q}^\alpha+\]
\[+\frac{1}{2}\left(\frac{\partial\mu_{\alpha\beta}}{\partial q^{\gamma}}-\frac{\partial\mu_{\gamma\beta}}{\partial q^{\alpha}}-\frac{\partial\mu_{\gamma\alpha}}{\partial q^{\beta}}\right)\,\dot{q}^\alpha \dot{q}^{\beta}\,.\]
These equations are the Lagrange equations for a natural system \cite[formula (1.6)]{30}.

	\subsubsection{Equations of trajectorial mechanics as equations of motion.}
	
	Let us cast equation \eqref{2.21} into a purely geometric form. To do this, we divide equation \eqref{2.21} by $\sqrt{2T}$. We then obtain
\begin{equation} \label{2.16}
	\frac{dT}{dq}=\left(-\frac{\partial U }{\partial q^\alpha}-\frac{\partial P_\alpha}{\partial t}\right)\tau ^\alpha  - \frac{\sqrt{2T}}{2} \,\frac{\partial \mu _{\alpha \beta }}{\partial t}\,\tau ^\alpha \tau ^\beta\,.
\end{equation}
We shall treat the equations \eqref{2.20} and \eqref{2.16} as a system of ordinary differential equations—of the second  and first order, respectively—in which $q$ serves as the independent variable. The initial conditions encompass the kinetic energy, $s$ coordinates, and $s-1$ independent components of the tangent vector (given the normalization condition $\mu_{\alpha\beta}\tau^\alpha\tau^\beta=1$ for the tangent vector). Consequently, excluding the initial time instant, the total number of initial constants is $2s$. This corresponds to the number of initial conditions in Lagrangian mechanics ($s$ coordinates and $s$ velocities), though expressed here in terms of trajectory parametrization: $s$ coordinates, $s-1$ independent components of the tangent vector, and the kinetic energy.


In the important special case where a natural system is conservative, the external fields and the metric of the configuration space—which appear on the right-hand sides of the equations under discussion—do not depend on time. In this case, the equation \eqref{2.20} is sufficient to determine the entire trajectory. Once this is done, integrating \eqref{1.4} along the resulting curve completely solves the problem of determining the law of motion. The equation \eqref{2.16}, meanwhile, is not required for finding the trajectory, as it is automatically satisfied by virtue of the law of conservation of energy, $T+U=E=\text{const}$. Let us verify this conclusion. Indeed, if the external fields and the natural system itself remain unchanged, the equation \eqref{2.16} can be rewritten as follows: 
		\[\left(\frac{\partial T}{\partial q^\alpha}+\frac{\partial U}{\partial q^\alpha}\right)\tau^\alpha \equiv 0\,,\]
	which, obviously, follows from the condition $T=E-U$.

In the general case of a non-conservative system, the right-hand side of equations \eqref{2.20} and \eqref{2.16} also depends on time, and the number of unknowns exceeds the number of equations by one. However, if one adds to these equations the relation \eqref{1.10} or \eqref{1.4}—along with the definition of the refractive index given by \eqref{1.5}—then there are enough equations to obtain a complete solution using standard methods. Therefore, the aforementioned equations serve as the general equations of motion.

\subsubsection{Natural form of equations.} 

We shall demonstrate that the equations of trajectory mechanics \eqref{2.20} and \eqref{2.16}, taken together with the relation \eqref{1.10}, are, in fact, “natural” equations of motion—that is, they describe the state of a system in a natural way, namely, by means of the trajectory and the time elapsed along it. In other words, we shall prove that the solutions to the equations of trajectory mechanics provide a natural parametrization of both the trajectory and the time. 

To do this, we transform the equations of trajectory mechanics \eqref{2.20} and \eqref{2.16} into a form more convenient for solving them. We multiply equation \eqref{2.20} by $\mu^{\lambda\gamma}/2T$ and simplify:
\[\frac{d\tau^{\lambda}}{dq}=-\frac{\mu^{\lambda\alpha}-\tau^{\alpha}\tau^{\lambda}}{2T}\left(\frac{\partial U}{\partial q^{\alpha}}+\frac{\partial P_{\alpha}}{\partial t}\right)+\frac{\mu^{\lambda\gamma}}{\sqrt{2T}}\left(\frac{\partial P_{\alpha}}{\partial q^{\gamma}}-\frac{\partial P_{\gamma}}{\partial q^{\alpha}}-\frac{\partial \mu_{\gamma\alpha}}{\partial t}\right)\,\tau^{\alpha}+\]
		\begin{equation}\label{2.19}
	+\frac{\mu^{\lambda\gamma}}{2}\left(\frac{\partial\mu_{\alpha\beta}}{\partial q^{\gamma}}-\frac{\partial\mu_{\gamma\beta}}{\partial q^{\alpha}}-\frac{\partial\mu_{\gamma\alpha}}{\partial q^{\beta}}\right)\,\tau^{\alpha} \tau^{\beta}+\frac{1}{2\sqrt{2T}}\,\frac{\partial \mu_{\alpha\beta}}{\partial t}\,\tau^{\alpha}\tau^{\beta} \tau^{\lambda}\,.
\end{equation}

Next, in equation \eqref{2.19}, we take into account that $1/\sqrt{2T} = dt/dq$ and replace the components of the tangent vector $\tau^\alpha$ in $dq^\alpha/dq$. 
\[\frac{d^{\,2}q^{\lambda}}{dq^{2}}=-\mu^{\lambda\alpha}\left(\frac{\partial U}{\partial q^{\alpha}}+\frac{\partial P_{\alpha}}{\partial t}\right)\left(\frac{dt}{dq}\right)^2+\mu^{\lambda\gamma}\left(\frac{\partial P_{\alpha}}{\partial q^{\gamma}}-\frac{\partial P_{\gamma}}{\partial q^{\alpha}}-\frac{\partial \mu_{\gamma\alpha}}{\partial t}\right)\,\frac{dq^\alpha}{dq}\frac{dt}{dq}+\]
			\[+\frac{\mu^{\lambda\gamma}}{2}\left(\frac{\partial\mu_{\alpha\beta}}{\partial q^{\gamma}}-\frac{\partial\mu_{\gamma\beta}}{\partial q^{\alpha}}-\frac{\partial\mu_{\gamma\alpha}}{\partial q^{\beta}}\right)\,\frac{dq^\alpha}{dq} \frac{dq^\beta}{dq}+\left(\frac{\partial U}{\partial q^{\alpha}}+\frac{\partial P_{\alpha}}{\partial t}\right)\frac{dq^\alpha}{dq} \frac{dq^\lambda}{dq}\left(\frac{dt}{dq}\right)^2+\]
	\begin{equation}\label{2.1}
	+\frac{1}{2}\,\frac{\partial \mu_{\alpha\beta}}{\partial t}\,\frac{dq^\alpha}{dq} \frac{dq^\beta}{dq} \frac{dq^\lambda}{dq}\frac{dt}{dq}\,.
\end{equation}

In equation \eqref{2.16} we define kinetic energy as
\[	T=\frac{1}{2}\,\left(\frac{dt}{dq}\right)^{-2}\,,\]
which follows from \eqref{1.10}. Next, we differentiate the left-hand side with respect to $q$, multiply both sides of the resulting equation by $-\left(dt/dq\right)^3$, and replace the components of the tangent vector $\tau^\alpha$ with $dq^\alpha/dq$. This yields the following equation
\begin{equation} \label{2.22}
	\frac{d^2t}{dq^2}= \frac{1}{2}\,\frac{\partial \mu _{\alpha \beta }}{\partial t}\,\frac{dq^\alpha}{dq} \frac{dq^\beta}{dq}\left(\frac{dt}{dq}\right)^2+\left(\frac{\partial U }{\partial q^\alpha}+\frac{\partial P_\alpha}{\partial t}\right)\frac{dq^\alpha}{dq}\left(\frac{dt}{dq}\right)^3 \,.
\end{equation}

In the resulting system of equations for Jacobi mechanics, \eqref{2.1} and \eqref{2.22}, the trajectory and the law of motion along that trajectory are interdependent. These equations are analogous to the Lagrange equations, but more complex; specifically, there is one more equation here than in the Lagrange formulation. Clearly, the $s+1$ equations \eqref{2.1} and \eqref{2.22} are of the second order. At present, it is not known how to solve them. In principle, however, the simultaneous integration of equations \eqref{2.1} and \eqref{2.22} allows for the determination of the functions that constitute their solution. These functions depend on $q$ and $2s$ initial constants: $q^\alpha = q^\alpha(q, T_0, q^\alpha_0, \tau^\alpha_0)$ and $t = t(q, T_0, q^\alpha_0, \tau^\alpha_0)$. Along with this solution, the kinetic energy $T(q)$ is also determined.

	\subsubsection*{Discussion}

Thus, the consequences of Jacobi’s principle of stationary action, with respect to variations in the law of motion (expressed in its natural form), are given by equations \eqref{2.21} and \eqref{2.20}. Equation \eqref{1.4} also follows from Jacobi’s action, provided one assumes stationarity with respect to variations in energy as well \cite[formula (44.12)]{15}. All these equations were also derived in \cite{30} from the Lagrange equations for natural systems, independently of Jacobi’s principle.

As demonstrated in Section 5, the converse also holds: the Lagrange equations can be rigorously derived from the consequences of the theorem under discussion, taken together with equation \eqref{1.4}. Consequently, one must conclude that the Lagrange equations are equivalent to the combination of the law governing the change in kinetic energy, the equation of the trajectory, and the speed of motion along that trajectory. Furthermore, since the consequences of both forms of the principle of stationary action are identical, the principles themselves are equivalent, as they lead to the same conclusions. Therefore, Jacobi’s form of the principle of stationary action—contrary to the earlier, more limited interpretation (where the principle was applied only to conservative mechanical systems)—should be understood as a complete geometric reformulation of the Hamilton-Ostrogradsky principle.

The physical significance of equations \eqref{1.4}, \eqref{2.21}, and \eqref{2.20} is as follows. Equation \eqref{1.4} asserts that, in configuration space, a natural system moves along its trajectory with a speed $n$.
Equation \eqref{2.21} describes the change in the system's kinetic energy. Equation \eqref{2.20} simultaneously serves as the equation for the trajectory of a natural system in configuration space. It is a second-order equation that describes the evolution of the tangent vector, depending on initial conditions, external fields, and the properties of the system itself. For instance, if a natural system is at a given point, and both the tangent vector and the total energy at that point are known, then this equation determines the tangent vector at an adjacent point at the next instant in time. In other words, this relation dictates the local geometric shape of the trajectory. It is well known that the Jacobi principle governs the trajectory of a mechanical system. In the traditional formulation of this principle, it was not clear how to determine the trajectory of a mechanical system within a time-varying external field. The modified Jacobi principle, however, addresses this issue through equation \eqref{2.20}.

	The aforementioned equations can be reduced to a natural form: the system \eqref{2.1}, \eqref{2.22}. In the Jacobi method, the solution to the equations of motion is understood as the "natural parametrization" of the trajectory \cite[ch. 3, p. 2.1]{33}, along with the corresponding parametrization of time: $q^\alpha=q^\alpha(q)$ and $t=t(q)$. This conclusion is consistent with the Jacobi principle discussed in this paper. Therefore, Jacobi’s trajectory-based mechanics constitutes a self-consistent theory.

Thus, in light of the above, the theorem regarding the stationarity of the Jacobi total action is proven, and the Jacobi action and the Hamilton–Ostrogradsky action are shown to be equivalent.
The conclusion that these two types of action — \eqref{1.1} and \eqref{1.8} — are identical is significant. It implies that, by taking the principle of stationary action in the Jacobi form as a foundation — analogous to the principle in the Hamilton–Ostrogradsky form — one can construct the entire analytical mechanics of natural systems.

Why, then, was the necessity of accounting for the variation of time in the Jacobi action not noticed earlier? The reason is that, for conservative systems, the integral term involving $\delta t$ in the general variation \eqref{4.8} vanishes when evaluated along the solutions to the equations of motion—even though, in reality, $\delta t \neq 0$. This specific case, in which the variation of the action with respect to time disappears, was inadvertently generalized to non-conservative systems. In other words, the importance of varying the Jacobi action with respect to time for non-conservative systems remained obscured. This is precisely why the requirement for an additional variational treatment of time long went unnoticed. However, in the general case, restricting the Jacobi principle solely to conservative systems is unjustified.

At the end of Section 4, the variation of the Jacobi action—interpreted as a function of the limits of integration—was calculated. Routh demonstrated \cite[ch. X, p. 442]{27} that, when coordinates and time are varied independently of one another, the variation of the Hamilton–Ostrogradsky action \eqref{1.1} is given by
\begin{equation}
\delta S=\left[L\delta t+\frac{\partial L}{\partial \dot{q}^\alpha}\left(\delta q^\alpha-\dot{q}^\alpha \delta t\right)\right]\bigg|^{t_2}_{t_1}+\int\limits^{t_2}_{t_1} \left(\frac{d}{d t}\frac{\partial L}{\partial \dot{q}^\alpha}-\frac{\partial L}{\partial q^\alpha}\right)\,\left(\delta q^\alpha-\dot{q}^\alpha \delta t\right)dt\,.
\end{equation}
 Comparing the first term of this expression with \eqref{6.21}, we see that they are identical, up to a change in parameterization ($t$ replaced by $q$). Therefore, the meaning of the variations with respect to coordinates and time remains the same, regardless of the specific form of the action; only the parameterization of these variations may differ.

Consequently, the initial assumption—that the operator for variations with respect to coordinates and time remains invariant under a change of the action—is justified.

Let us compare these results with the conclusions reached by other researchers.

In the Jacobi action, the integrand depends on both coordinates and time. For this reason, the maining of the Jacobi integral for non-conservative systems is difficult to grasp without referencing the "natural" parameterization used for the solutions to the equations of motion. However, this crucial issue is not addressed in the cited literature.

In the monograph \cite[Chapter VII, Section 4, Subsection 3]{23}, it is also shown that Lagrange's equations can be derived from the principle of least action in the Jacobi form, and that Jacobi's principle follows from the Hamilton-Ostrogradsky principle. However, the proof is provided only for conservative systems, unlike in the present work.

\begin{flushleft}
{\bf{Conclusion.}}
\end{flushleft}

This study focuses on the mechanics of non-conservative natural systems, based on the Jacobi total action functional.

In the course of the research, the following results have been obtained for such mechanical systems:

a) A theorem has been formulated and proven stating that the Jacobi total action is stationary not only under independent variations of the coordinates but also under an independent variation of time. These variations differ from those previously known in that they depend on the natural parametrization of the trajectory and time—rather than on a temporal parameter—and, furthermore, the variation of time vanishes at the endpoints of the integration path. Consequently, a modification of the well-known variational method has been achieved, and the "trajectory-based" formulation of one of the three fundamental variational principles of mechanics has been completed; 

b) it has been established that the well-known equations for the rate of change of kinetic energy \eqref{2.21} and for the trajectory \eqref{2.20} are rigorous consequences of the principle under discussion; this provides a justification for a geometric approach to the study of natural systems; 

c) it has been shown that the Lagrange equations follow from these equations and the equality \eqref{1.4}; 

d) it has been found that the meaning of the variations of coordinates and time is the same across various forms of the principle of stationary action; this confirms the invariance of the variation operator; 

e) "natural" equations of motion \eqref{2.1}, \eqref{2.22} have been derived as a system of $s+1$ second-order equations, the solutions to which possess a natural parametrization. 

As a result of the work performed, the following sequential conclusions can be drawn. The principles of stationary action in all three forms are equivalent and interchangeable. Therefore, the Jacobi method can be applied to non-conservative natural systems. Consequently, this method serves as a robust and independent geometric formulation of analytical mechanics. Thus, the objective of the research has been achieved. In summary, it can be stated that the principle of stationary action in the Jacobi form—as examined in this study—along with the newly derived trajectory equations, provide a reliable foundation for further research.
\footnote{Please cite this article in press as:\\
Voytik V. V. Jacobi’s principle of stationary full action and its implications, Vestn. Samar.
Gos. Tekhn. Univ., Ser. Fiz.-Mat. Nauki [J. Samara State Tech. Univ., Ser. Phys. Math. Sci.],
2026, vol. 30, no. 3, pp. 598–619. EDN: CLCXLV. DOI: 10.14498/vsgtu2312 (In Russian).}


\newpage
\begin{thebibliography}{99}
 %
\bibitem{18}
{Polak L. S.} Variatsionnye printsipy mekhaniki, ikh razvitie i primenenie v fizike  \rm[Variational principles of mechanics: Their development and applications in physics](Russian)
2016. 600 p. Moscow: Librokom. In Russian.\url{https://zbmath.org/0101.41402}

\bibitem{34}
Variatsionnye printsipy mekhaniki. Sb. statei klassikov nauki \rm[Variational principles of mechanics. Collection of articles by classics of science] ed. Polak L. S. 1959. 932 p. Moscow. Gos. izd. fiz.-mat. lit. In Russian. \url{https://zbmath.org/0087.17002}

\bibitem{1}
{Jacobi C. G. J.} Jacobi’s lectures on dynamics. Delivered at the University of K{o}nigsberg in the winter semester 1842–1843 and according to the notes prepared by C. W. Borchardt. Vol.51 / Texts and Readings in Mathematics/ ed. A. Clebsch. New Delhi: Hindustan Book Agency, 2009. 339 p. 
\url{https://doi.org/10.1007/978-93-86279-62-0}.

\bibitem{77}
{Hertz H.} The principles of mechanics: presented in a new form. Authorized English translation by E. Jones and J. T. Walley. 1899, p. 276. London, Macmillan and Co.  \url{https://zbmath.org/30.0617.07}

\bibitem{44}
{Abraham R. H., Marsden J. E.} Foundations of mechanics. With the assistance of Tudor Ratiu and Richard Cushman. 2nd ed. 1978. The Benjamin/Cummings Publishing Company Inc. Reading, Massachusetts. 806 p. \url{https://zbmath.org/0393.70001}
  
\bibitem{8}
{Arnold V. I.} Mathematical methods of classical mechanics. Translated by K. Vogtman and A. Weinstein. 1989. Graduate Texts in Mathematics. vol. 60. New York - Heidelberg - Berlin: Springer-Verlag. 462 p.\url{https://zbmath.org/0386.70001}, \url{https://doi.org/10.1007/978-1-4757-2063-1}

\bibitem{37}
{Arnold V. I., Kozlov V. V., Neishtadt A. I.} Mathematical Aspects of Classical and Celestial Mechanics. Transl. from the Russian by E. Khukhro. 3rd ed. Encyclopaedia of Mathematical Sciences. vol. 3. 2006. 518 p. Berlin, Heidelberg: Springer-Verlag. 
\url{https://doi.org/10.1007/978-3-540-48926-9}. \url{https://zbmath.org/1105.70002}

\bibitem{19}
 {Basdevant J.-L.} Variational Principles in Physics. 2023. Cham: Springer. 242 p. 
\url{https://zbmath.org/7674126}, \url{https://doi.org/10.1007/978-3-031-21692-3}

\bibitem{23}
 {Berezkin E.N.} Kurs teoreticheskoj mekhaniki [Course of theoretical mechanics]. 1974. 647 p. Moscow. Moscow University Press. In Russian

\bibitem{16}
 {Bir\'o T. S.} Variational Principles in Physics. From Classical to Quantum Realm. Translated from the Hungarian //SpringerBriefs in Physics. 2023. Cham: Springer.  112 p. 
\url{https://zbmath.org/7729624},
 \url{https://doi.org/10.1007/978-3-031-27876-1}

\bibitem{11}
 {Bolotin S. V., Karapetyan A. V., Kugushev E. I., Treschev D. V.} Teoreticheskaya mekhanika [Theoretical mechanics: textbook for students of higher education institutions]. 2010. Moscow: Izd. centr «Akademiya». 432 p. In Russian

\bibitem{29}
  {Bukhgolts N.N.} Osnovnoj kurs teoreticheskoj mekhaniki [Basic Course of Theoretical Mechanics]. Chast' vtoraya. Dinamika sistemy material'nyh tochek [Part Two. Dynamics of a System of Material Points]. 1966. Moscow: Nauka, Glav. red. fiz.-mat. lit. 332 p.  In Russian

\bibitem{36}
 {Vilke V. G.} Teoreticheskaya mekhanika: Uchebnik [Theoretical Mechanics: Textbook]. 1998. 272 p. Moscow: Moscow State University. In Russian

\bibitem{32}
 {Gantmacher F.} Lectures in analytical mechanics. Translated from the Russian by Yankovsky G.
 1970. Moscow: Mir Publishers. 264 p. \url{https://zbmath.org/0212.56701}

\bibitem{10}
 {Goldstein H., Poole Ch., Safko J.} Classical mechanics. 2002. 3rd ed. Addison Wesley. San Francisco, Sansome St., 1301. 638 p. \url{https://doi.org/10.1119/1.1484149}

\bibitem{7}
 {Golubev Yu. F.} Osnovy teoreticheskoj mekhaniki. Uchebnik [Fundamentals of Theoretical Mechanics: Textbook]. 2000. 719 c. Moscow State University. Moscow
\url{https://zbmath.org/0776.70002}. In Russian

\bibitem{20}
{Cline D.} Variational Principles in Classical Mechanics. 2021. 3rd ed. University of Rochester River Campus Libraries. Rochester, NY 14627. 542 p.

\bibitem{15}
 {Landau L. D., Lifshits E. M.} Mechanics. Translated by J. B. Sykes and J. S. Bell.
1976. 3rd ed. Course of Theoretical Physics. 1. 165 p.
 \url{https://zbmath.org/3181761}. Pergamon Press. Oxford-London-New York-Paris
\url{https://doi.org/10.1016/C2009-0-25569-3}

\bibitem{17}
{Lanczos~C.} The Variational Principles of Mechanics. Mathematical Expositions. 4th ed. 1970.
vol. 4. 418 p. \url{https://zbmath.org/3407221}. University of Toronto Press. Toronto
\url{https://doi.org/10.3138/9781487583057}

\bibitem{13}
{Levi-Civita T., Amaldi U.} A Course of Theoretical Mechanics. Vol. 2. Dynamics of systems with a finite number of degrees of freedom. Part two. 1951. 556 p. IL. Metelitsyn I. I.
Moscow. In Russian

\bibitem{26}
{Lurie A.I.} Analytical Mechanics. Translated from the Russian. With a foreword by V. A. Palmov and a preface by A. Belyaev. Foundations of Engineering Mechanics. 2002. Springer. Berlin, Heidelberg. 864 p.
\url{https://zbmath.org/1903882}, 
\url{https://doi.org/10.1007/978-3-540-45677-3}

\bibitem{31}
{Markeev A.P.} Teoreticheskaya mekhanika. Uchebnoe posobie [Theoretical Mechanics: Textbook].
1990. 415. Nauka. Moscow. \url{https://zbmath.org/0747.70002}.In Russian

\bibitem{12}
{Marsden J. E., Ratiu T. S.} Introduction to Mechanics and Symmetry: A Basic Exposition of Classical Mechanical Systems. vol 17. Texts in Applied Mathematics. 2nd ed. 1999. Springer New York. NY. 586 p. \url{https://doi.org/10.1007/978-0-387-21792-5}. \url{https://zbmath.org/0933.70003}

\bibitem{63}
{Novikov S. P., Fomenko A. T.} Basic elements of differential geometry and topology. Transl. from the Russian by M. V. Tsaplina. Mathematics and Its Applications. Soviet Series
vol. 60. 1990. Kluwer Academic Publishers. Dordrecht etc. 490 p.
\url{https://zbmath.org/0711.53001}

\bibitem{6}
{Nolting W.} Theoretical physics 2. Analytical mechanics. 2016.\url{https://zbmath.org/1339.70005}. Springer. Cham. 358 p. \url{https://doi.org/10.1007/978-3-319-40129-4}


\bibitem{28}
{Olkhovsky I. I.} Kurs teoreticheskoj mekhaniki dlya fizikov [Course of theoretical mechanics for physicists]. 1978. Moscow State University. Moscow. In Russian. 575 p.

\bibitem{21}
{Petkevich V.V.} Teoreticheskaya mekhanika: Uchebnoe posobie [Theoretical Mechanics: A Textbook]
 1981. Nauka. Gl. red. phys.-mat. lit. Moscow. In Russian. 496 p.

\bibitem{3}
{Polyakhov N. N., Zegzhda S. A., Yushkov M. P.} Teoreticheskaya mekhanika: uchebnik dlya akademicheskogo bakalavriata [Theoretical Mechanics: A Textbook for Academic Bachelor's Degree] Bakalavr. Akademicheskij kurs [Bachelor's degree. Academic course]. 2015. 592. Yurait. Tovstik P.E. Moscow. In Russian

\bibitem{2}
 {Suslov G. K.} Osnovy analiticheskoj mekhaniki: uchebnik [Fundamentals of Analytical Mechanics: Textbook]. 2020. 672 p. Fiziko - matematicheskoe nasledie: fizika (mekhanika) [Physical and mathematical heritage: physics (mechanics)]. LENAND. Buchgolts N. N., Goltsman V. K. Moscow. In Russian

\bibitem{46}
{Bolsinov A. V., Kozlov A. V., Fomenko A. T.} The Maupertuis principle and geodesic flows on the sphere arising from integrable cases in the dynamics of a rigid body Russian Math. Surveys, 1995. vol 50. issue 3. pp. 473--501.
\url{https://doi.org/10.1070/RM1995v050n03ABEH002100}
\url{https://www.webofscience.com/wos/woscc/full-record/WOS:A1995TT98400001}

\bibitem{48}
{Tsiganov A. V.} Equivalent Integrable Metrics on the Sphere with Quartic Invariants. SIGMA, 2022. vol 18, 094,
\url{https://zbmath.org/1514.37077}
\url{https://doi.org/10.3842/SIGMA.2022.094}

\bibitem{49} 
{Agapov S.V., Tursunov M.M.} On the Rational Integrals of Two-Dimensional Natural Systems //\textit{Siberian Math. J.} 2023. vol. 64. № 4. p. 787--795
\url{https://doi.org/10.1134/S0037446623040018}

\bibitem{50}
{Tsiganov A. V.} Integrable Systems on a Sphere, an Ellipsoid and a Hyperboloid. Regul. Chaotic Dyn. 2023. vol 28. issue 6. pp. 805--821
\url{http://mi.mathnet.ru/rcd1234}
\url{https://doi.org/10.1134/S1560354723520088}

\bibitem{51} 
{Fomenko A. T., Vedyushkina  V. V.} Billiards and integrable systems //Russian Math. Surveys. 2023. vol. 78. iss. 5. p. 881--954
\url{https://doi.org/10.4213/rm10100e}
\url{https://gateway.webofknowledge.com/gateway/Gateway.cgi?GWVersion=2&SrcApp=Publons&SrcAuth=Publons_CEL&DestLinkType=FullRecord&DestApp=WOS_CPL&KeyUT=001184355800003}
\url{https://www.scopus.com/record/display.url?origin=inward&eid=2-s2.0-85191510714}

\bibitem{52}
{Agapov S. V., Mironov  A. E.} Finite-Gap Potentials and Integrable Geodesic Equations on a 2-Surface //Proc. Steklov Inst. Math. 2024. vol.327, p. 1–11 
\url{https://doi.org/10.1134/S0081543824060014},
\url{https://www.scopus.com/record/display.url?origin=inward&eid=2-s2.0-105001525773}

\bibitem{47}
 {Henheik J.} Deformational rigidity of integrable metrics on the torus//Ergodic Theory and Dynamical Systems. 2024. vol 45. issue 2. p. 467--503
\url{https://doi.org/10.1017/etds.2024.48}

\bibitem{53}
{Kress J., Sch\"obel K., Vollmer A.} Algebraic Conditions for Conformal Superintegrability in Arbitrary Dimension//Commun. Math. Phys. 2024. vol. 405. issue 92.
\url{https://doi.org/10.1007/s00220-023-04872-w}

\bibitem{54}
{Agapov S.} Local high-degree polynomial integrals of geodesic flows and the generalized hodograph method//Journal of Geometry and Physics. 2025. vol. 217. p. 105629.
\url{https://doi.org/10.1016/j.geomphys.2025.105629}

\bibitem{55}
{Matveev V.S.} Real Analyticity of 2-Dimensional Superintegrable Metrics and Solution of Two Bolsinov – Kozlov – Fomenko Conjectures//Regul. Chaot. Dyn. 2025. vol. 30. issue 4. p. 677–687
\url{https://doi.org/10.1134/S1560354725040148}

\bibitem{38}
{Dobrokhotov S.Y., Nosikov I.A., Tolchennikov A.A.} The Maupertuis–Jacobi Principle and the Fermat Variational Principle in the Problem of Short-Wavelength Asymptotics of the Solution of the Helmholtz Equation with a Localized Source// Comput. Math. and Math. Phys. 2025. vol. 65, iss. 4. p. 739–753. 
\url{https://doi.org/10.1134/S0965542525700010}

\bibitem{56}
{Kozlov V.~V.} Maupertuis's principle for systems with Lagrangians linear in velocities // \textit{Sb. Math.} 2025. vol. 216. iss. 5. p. 714--722
\url{https://doi.org/10.4213/sm10185e}

\bibitem{35}
{Courant R., Hilbert D.} Methods of Mathematical Physics. vol I. Reprint of the 1st Engl. ed. 
1989. New York etc.: John Wiley and Sons. 560 p.
\url{https://zbmath.org/0729.00007}

\bibitem{30} 
{Voytik V. V.} Trajectory equations for a non-conservative natural system //\textit{Vestn. Tomsk. Gos. Univ. Matematika i mekhanika} [Tomsk State University Journal of Mathematics and Mechanics], 2025. no 95, p. 72--80. \url{https://doi.org/10.17223/19988621/95/7} (in Russian),
\url{https://doi.org/10.48550/arXiv.2410.05682} 

\bibitem{25}
{Landau L. D., Lifshitz E. M.} The classical theory of fields. 2nd ed. Paris-Frankfurt: Pergamon Press. 1962. 404 p.
\url{https://zbmath.org/0178.28704}
\url{https://doi.org/10.1002/zamm.19630430611}

\bibitem{4}
{Sardanashvili G. A.} Sovremennye metody teorii polya [Modern Methods in Field Theory]. Vol 2. Geometriya i klassicheskaya mekhanika [Geometry and Classical Mechanics]. 2024. Moscow: LKI.
168 p. In Russian

\bibitem{22}
{Kittel Ch.} Introduction to Solid State Physics. 8th ed. 2005. New York: John Wiley and Sons Inc.
704 p.
\url{https://zbmath.org/0052.45506}
\url{https://doi.org/10.1107/S0365110X54000448}

\bibitem{33}
{Budak B.M., Fomin S.V.} Multiple Integrals, Field Theory And Series. Transl. from the Russian by V. M. Volosov. An Advanced Course in Higher Mathematics. 1973. Moscow: Mir Publishers. 640 p.
\url{https://zbmath.org/0267.00004}

\bibitem{27}
{Routh E. J. and etc.} The advanced part of a treatise on the dynamics of a system of rigid bodies: being part II. of a treatise on the whole subject, with numerous examples. 5th ed.
1892. 452 p. London, New York: Macmillan and Co.
\url{https://zbmath.org/24.0812.10}


 \end{thebibliography}
\end {document}